\documentclass[twoside,twocolumn,9pt]{article}
\usepackage{extsizes}
\usepackage[super,sort&compress,comma]{natbib} 
\usepackage[version=3]{mhchem}
\usepackage[left=1.5cm, right=1.5cm, top=1.785cm, bottom=2.0cm]{geometry}
\usepackage{balance}
\usepackage{times,mathptmx}
\usepackage{sectsty} 
\usepackage{graphicx}
\usepackage{subfigure} 
\usepackage{enumitem}
\usepackage{url}
\usepackage{epsfig}
\usepackage{lastpage}
\usepackage[format=plain,justification=justified,singlelinecheck=false,font={stretch=1.125,small,sf},labelfont=bf,labelsep=space]{caption}
\usepackage{float}
\usepackage{fancyhdr}
\usepackage{fnpos}
\usepackage[english]{babel}
\addto{\captionsenglish}{
}
\usepackage{array}
\usepackage{droidsans}
\usepackage{charter}
\usepackage[T1]{fontenc}
\usepackage[usenames,dvipsnames]{xcolor}
\usepackage{setspace}
\usepackage[compact]{titlesec}
\usepackage{hyperref}
\usepackage{epstopdf}
\usepackage{chemformula} 
\usepackage{gensymb}
\usepackage{color}
\usepackage{soul}
\usepackage[utf8]{inputenc}
\usepackage{amsmath, amsfonts, mathtools, amssymb, amsthm, bm}
\usepackage{dcolumn}
\usepackage{tabularx}
\usepackage{here}
\usepackage{pslatex}          
\usepackage{textcomp}
\usepackage{multirow, makecell}
\usepackage{rotating} 
\usepackage{pifont}
\usepackage{lmodern}
\usepackage{microtype}
\usepackage{ulem} 

\definecolor{cream}{RGB}{222,217,201}

\begin{document}

\pagestyle{fancy}
\thispagestyle{plain}
\fancypagestyle{plain}{

\renewcommand{\headrulewidth}{0pt}
}

\makeFNbottom
\makeatletter
\renewcommand\LARGE{\@setfontsize\LARGE{15pt}{17}}
\renewcommand\Large{\@setfontsize\Large{12pt}{14}}
\renewcommand\large{\@setfontsize\large{10pt}{12}}
\renewcommand\footnotesize{\@setfontsize\footnotesize{7pt}{10}}
\makeatother

\renewcommand{\thefootnote}{\fnsymbol{footnote}}
\renewcommand\footnoterule{\vspace*{1pt}%
\color{cream}\hrule width 3.5in height 0.4pt \color{black}\vspace*{5pt}} 
\setcounter{secnumdepth}{5}

\makeatletter 
\renewcommand\@biblabel[1]{#1}            
\renewcommand\@makefntext[1]%
{\noindent\makebox[0pt][r]{\@thefnmark\,}#1}
\makeatother 
\renewcommand{\figurename}{\small{Fig.}~}
\sectionfont{\sffamily\Large}
\subsectionfont{\normalsize}
\subsubsectionfont{\bf}
\setstretch{1.125} 
\setlength{\skip\footins}{0.8cm}
\setlength{\footnotesep}{0.25cm}
\setlength{\jot}{10pt}
\titlespacing*{\section}{0pt}{4pt}{4pt}
\titlespacing*{\subsection}{0pt}{15pt}{1pt}

\fancyfoot{}
\fancyfoot[LO,RE]{\vspace{-7.1pt}\includegraphics[height=9pt]{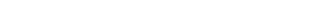}}
\fancyfoot[CO]{\vspace{-7.1pt}\hspace{13.2cm}\includegraphics{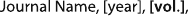}}
\fancyfoot[CE]{\vspace{-7.2pt}\hspace{-14.2cm}\includegraphics{head_foot/RF}}
\fancyfoot[RO]{\footnotesize{\sffamily{1--\pageref{LastPage} ~\textbar  \hspace{2pt}\thepage}}}
\fancyfoot[LE]{\footnotesize{\sffamily{\thepage~\textbar\hspace{3.45cm} 1--\pageref{LastPage}}}}
\fancyhead{}
\renewcommand{\headrulewidth}{0pt} 
\renewcommand{\footrulewidth}{0pt}
\setlength{\arrayrulewidth}{1pt}
\setlength{\columnsep}{6.5mm}
\setlength\bibsep{1pt}

\makeatletter 
\newlength{\figrulesep} 
\setlength{\figrulesep}{0.5\textfloatsep} 

\newcommand{\topfigrule}{\vspace*{-1pt}%
\noindent{\color{cream}\rule[-\figrulesep]{\columnwidth}{1.5pt}} }

\newcommand{\botfigrule}{\vspace*{-2pt}%
\noindent{\color{cream}\rule[\figrulesep]{\columnwidth}{1.5pt}} }

\newcommand{\dblfigrule}{\vspace*{-1pt}%
\noindent{\color{cream}\rule[-\figrulesep]{\textwidth}{1.5pt}} }

\makeatother

\twocolumn[
  \begin{@twocolumnfalse}
{\includegraphics[height=30pt]{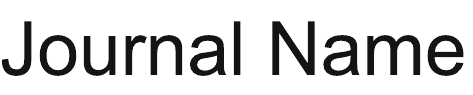}\hfill\raisebox{0pt}[0pt][0pt]{\includegraphics[height=55pt]{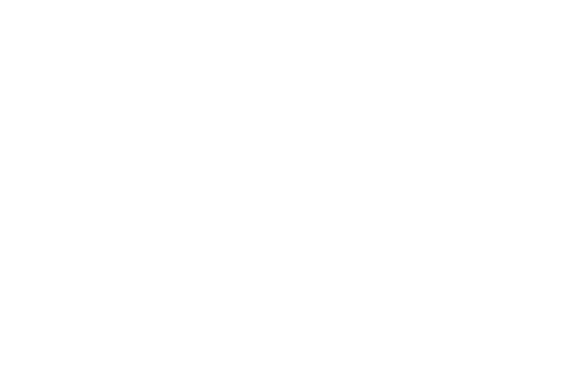}}\\[1ex]
\includegraphics[width=18.5cm]{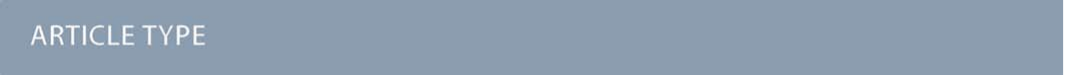}}\par
\vspace{1em}
\sffamily
\begin{tabular}{m{4.5cm} p{13.5cm} }

\includegraphics{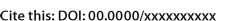} & \noindent\LARGE{\textbf{Prediction of induced magnetism in 2D Ti$_{{\rm2}}$C based MXene by manipulating the mixed surface functionalization and metal substitution computed by xTB model Hamiltonian of the DFTB method $^\dag$}} \\
\vspace{0.3cm} & \vspace{0.3cm} \\

 & \noindent\large{Taoufik Sakhraoui,$^{\ast}$\textit{$^{a}$} Frantisek Karlicky,\textit{$^{a}$}} \\

\includegraphics{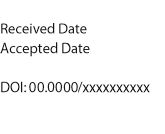} & \noindent\normalsize{We employed the recently developed density functional tight binding (DFTB) method's Hamiltonian, GFN1-xTB, for modeling the mixed termination in Ti$_{{\rm2}}$C MXene, namely three types of termination by combining -O and -OH, -O and -F, as well as -F and -OH. We demonstrated that the approach yields reliable predictions for electronic and magnetic properties of such MXenes. The first highlighted result is that the mixed surface functionalization in Ti$_{{\rm2}}$CA$_{{\rm x}}$B$_{{\rm y}}$ MXene induces spin polarization with diverse magnetic alignments, including ferromagnetism and two types of antiferromagnetism. We further identified the magnetic alignment for the investigated MXene in terms of the compositions of the terminal groups. Moreover, the effect of the transition metal (Ti) substituted by the Sc atom on the electronic and magnetic properties were also investigated. We found that the studied systems maintain the magnetism and the metallic characteristics. A magnetic transition from antiferromagnetic (AFM) to ferrimagnetic (FiM) ordering was found for  ScTi$_{{\rm15}}$C$_{{\rm8}}$F$_{{\rm8}}$(OH)$_{{\rm8}}$ and ScTi$_{{\rm15}}$C$_{{\rm8}}$F$_{{\rm12}}$(OH)$_{{\rm4}}$ compounds. Finally, we proved that incorporating the Sc atom into the lattice of Ti$_{{\rm2}}$CO$_{{\rm2}}$ and mixed surface termination Ti$_{{\rm2}}$CA$_{{\rm x}}$B$_{{\rm y}}$ is an effective strategy to induce magnetism. Our study may provide a new potential application for designing MXene-based spintronics.} \\
\end{tabular}
 \end{@twocolumnfalse} \vspace{0.6cm}
  ]

\renewcommand*\rmdefault{bch}\normalfont\upshape
\rmfamily
\section*{}
\vspace{-1cm}


\footnotetext{\textit{$^{a}$~Department of Physics, Faculty of Science, University of Ostrava, 701 03 Ostrava, Czech Republic}}
\footnotetext{\textit{$^{\ast}$~E-mail: taoufik.sakhraoui@osu.cz}}

\footnotetext{\dag~Electronic Supplementary Information (ESI) available: [details of any supplementary information available should be included here]. See DOI: ....}



\section{\label{sec1}Introduction}
In recent years, MXenes M$_{{\rm n}}$X$_{{\rm n-1}}$T$_{{\rm x}}$ ($1\le {{\rm n}} \le4$ and ${{\rm x}} \le 2$), a family of transition-metal carbide and nitride materials, consist of transition metals (M) including Sc, Ti, V, Cr, Mn, Y, Zr, Nb, Mo, Hf, Ta, carbon or nitrogen (X), and the surface termination (T), devoted great attention \cite{Anasori2022, Anasori2017, Zhang2019, Hope2016}. MXenes \cite{Anasori2017, YZhang2019}, as two-dimensional (2D) materials, have been experimentally synthesized \cite{Naguib2011} by using various experimental techniques, such as simple oxidation \cite{Johari2011}, nuclear magnetic resonance (NMR) spectroscopy \cite{Harris2015, Hope2016}, neutron diffraction \cite{Huan-Huan2016}, and STEM-EELS \cite{Karlsson2015} and theoretically predicted \cite{Khazaei2013, Changying2019, Ketolainen2022, monACSOmega, Dubecky2023, Kumar2023}. They have attracted great interest due to their excellent mechanical, electrical, optical, and electrochemical properties \cite{Sarikurt2018, Hu2017, Guo2016}, which have potential applications in various fields \cite{Hu2013, Gao2016, Khan2017, Yaqing2019} such as sensors, catalysis, energy storage, and nanoelectronics. Among MXenes, Ti$_{{\rm2}}$C is one of the most widely investigated, it was prepared by etching the Ti$_{{\rm2}}$AlC MAX phase \cite{Harith2023}. The arrangement between Ti and C atoms in Ti$_{{\rm2}}$C forms the honeycomb structure where carbon is sandwiched by two titanium. Moreover, during the etching process, Ti$_{{\rm2}}$C may be functionalized by F, O or OH forming Ti$_{{\rm2}}$CT$_{{\rm2}}$ (T = F, O, OH) \cite{Zhang2019, Khazaei2013, Zha2015, Hope2016, Ibragimova2016}. In these structures, the C atoms are still sandwiched between the Ti atoms, and the T atoms on both terminals occupy the top site of bottom titanium on both sides \cite{Zhang2019, Khazaei2013, Zha2015}. We note that previous literature reveals that surface terminal groups such as -Cl, -Br, -I, -S, -Te, -NH, -H, -BrI, and -ClBrI \cite{Li_2021, Kawai_2023, pang_2019, li_2019, vladislav_2020} were identified in several MXenes too.
It was recently predicted that the bare Ti$_2$C MXene prefers a magnetic ground state that corresponds to antiferromagnetic (AFM) coupled to ferromagnetic (FM) layers \cite{Garcia2023}. Whereas, the magnetic ground state of Ti$_2$C can be switched from AFM to nonmagnetic (NM) state by full surface termination by O, F and OH \cite{Bae2021, Jin2019, Enyashin2013}.
On the other hand, the control of the surface functionalization by O, OH, and F in MXenes may allow the control of the material properties. From experimental point of view, it has been shown that Ti$_{{\rm3}}$C$_{{\rm2}}$ showed a mixture of terminations by F, O, and OH \cite{Hope2016, Sang2016, Ibragimova2016} and that the composition depends on etching process, i.e., O functionalization dominates when using LiF and HCl \cite{Hope2016, Sang2016} and F functionalization dominates when using HF \cite{Hope2016}. MXenes naturally lose their inversion symmetry upon mixed surface functionalization, thereby opening up new design opportunities for controlling their physical properties. However, a proper understanding of the effect of the mixed functionalization of the Ti$_{{\rm2}}$C surface on the electronic and magnetic properties is still missing. Furthermore, to the extent of our knowledge, computational \cite{Ibragimova2016, Bagheri2021, Hu2018} and experimental \cite{Hope2016, Sang2016} studies did not treat magnetism in mixed terminated Ti$_{{\rm2}}$C. M. Bagheri {\it et al.} reported results for the Fermi surfaces of titanium carbide and nitride MXenes (Ti$_{{\rm3}}$C$_{{\rm2}}$, Ti$_{{\rm2}}$C, Ti$_{{\rm4}}$N$_{{\rm3}}$, and Ti$_{{\rm2}}$N) as a function of the mixed O and OH terminal groups \cite{Bagheri2021}. T. Schultz {\it et al.} reported a comparative study of a combination of ultraviolet photoemission spectroscopy (UPS), X-ray photoelectron spectroscopy (XPS), and inverse photoemission spectroscopy (IPES) measurements of Ti$_{{\rm3}}$C$_{{\rm2}}$T$_{{\rm x}}$ with density functional theory (DFT) calculations \cite{Schultz2019}.

The use of DFT for materials properties prediction is mostly applicable to small systems because of the high computational cost. This is a significant limitation because advances in MXenes require the study of larger systems, such as the influence of the concentration of vacancies and defects in the compound on the electronic and magnetic properties. Semiempirical methods, such as density functional tight binding (DFTB), may give a perfect and complementary platform \cite{dftb2016}. DFTB has been used to investigate numerous systems including solid state \cite{Haugk2000}, molecular and biological systems \cite{Elstner_2000}, titanium-based materials, \cite{monACSOmega, Yarkin2022} and 2D materials \cite{monACSOmega, monPCCP, manzano}. The applicability of DFTB method requires the availability of parameter sets for all pairs of atoms present in the system \cite{Nishimura2019, Hutama2021}. This approach has been already tested on the 2D materials \cite{Budiutama2023} and it was applied for defected Ti$_2$CO$_2$ MXene \cite{monACSOmega}. However, when compared with the exact eigenstates and eigenvalues from the exact calculations in finite Hubbard model \cite{Kocharian2006}, the tight binding case is a rather crude approximation. One can use exact calculations of local many body effects by incorporating the Hubbard or Heisenberg models in various structural geometries, as in refs. \cite{Auslaender2000, Sassetti1998, Lopez2008}. The Geometries, Frequencies, and Non-covalent interactions Tight Binding (GFN1-xTB) Hamiltonian presents a good alternative for electronic structure and magnetic properties calculations. GFN1-xTB Hamiltonian \cite{ref1_xTB} provides parameters for elements up to Z$\leq$86, and it was recently supported by DFTB+ code \cite{Aradi2020}. Moreover, the consistency between xTB and DFT functionals is already shown in recent publications \cite{ref4_xTB, Menzel2021}. The reader may refer to the following refs. \cite{ref1_xTB, ref2_xTB, ref3_xTB} for more details about xTB and comparison to other methods. 

In the present work, we employed GFN1-xTB version of the xTB Hamiltonian to investigate the electronic and magnetic properties of Ti$_{{\rm2}}$CA$_{{\rm x}}$B$_{{\rm y}}$ (A, B = O, F and OH and x and y = 0.5, 1.0 and 1.5; x + y = 2) and to study the effect of metal substitution in mixed surface termination Ti$_{{\rm2}}$C-based MXene. 

\section{Methods}
The xTB calculations presented in this study were performed using DFTB+ package\cite{Aradi2007, Aradi2020} with the implementation of the GFN1-xTB method \cite{ref1_xTB}. The GFN1-xTB Hamiltonian is well described in refs \cite{ref1_xTB, ref2_xTB} and \cite{ref4_xTB}. xTB combines high accuracy from DFT-based parameters with the reduced computational cost of standard tight-binding methods. This  technique is thus advantageous for describing large supercells. It has near DFT precision in electronic structure calculations while being significantly faster than DFT \cite{ref1_xTB, ref2_xTB, ref3_xTB, ref4_xTB, Vicent2021}. A conjugate gradient algorithm was used for geometry optimization imposing a maximum force of 10$^{{\rm -4}}$~au. The self-consistent calculations were considered to be converged when the difference in the total energy did not exceed 10$^{{\rm -3}}$~au. To obtain converged electron energy values, a Fermi distribution was used for the system at an electron temperature of 700 K. The total energy is calculated by integrating over the Brillouin zone with a Gamma-centered k-mesh of 6$\times$9$\times$1 Monkhorst-Pack \cite{monchorstpack} grid for the mixed terminal groups in the 2$\times$1$\times$1-Ti$_{{\rm2}}$C supercell (i.e., Ti$_{{\rm4}}$C$_{{\rm2}}$A$_{{\rm a}}$B$_{{\rm b}}$, where A and B are the F, O, and OH terminal groups and a + b = 4). The Ti substitution by Sc in Ti$_{{\rm2}}$CA$_{{\rm x}}$B$_{{\rm x}}$ is modeled by a 2$\times$2$\times$1 supercell and 1$\times$6$\times$1 grid. Finally, to study the effect of the Ti substitution by Sc in Ti$_{{\rm2}}$CO$_{{\rm2}}$, we used a 5$\times$5$\times$1 supercell (Ti$_{{\rm50}}$C$_{{\rm25}}$O$_{{\rm50}}$) and a $\Gamma$-point only setup. In order to prevent interaction with its own images, the out-of-plane bounding box was set to 20 \AA. Some additional DFT calculations were performed using Quantum Espresso (QE) package, see ESI for more details.

\section{Results and discussions}

\begin{figure}[H]
\begin{center}
\subfigure[ ] {\epsfig{figure=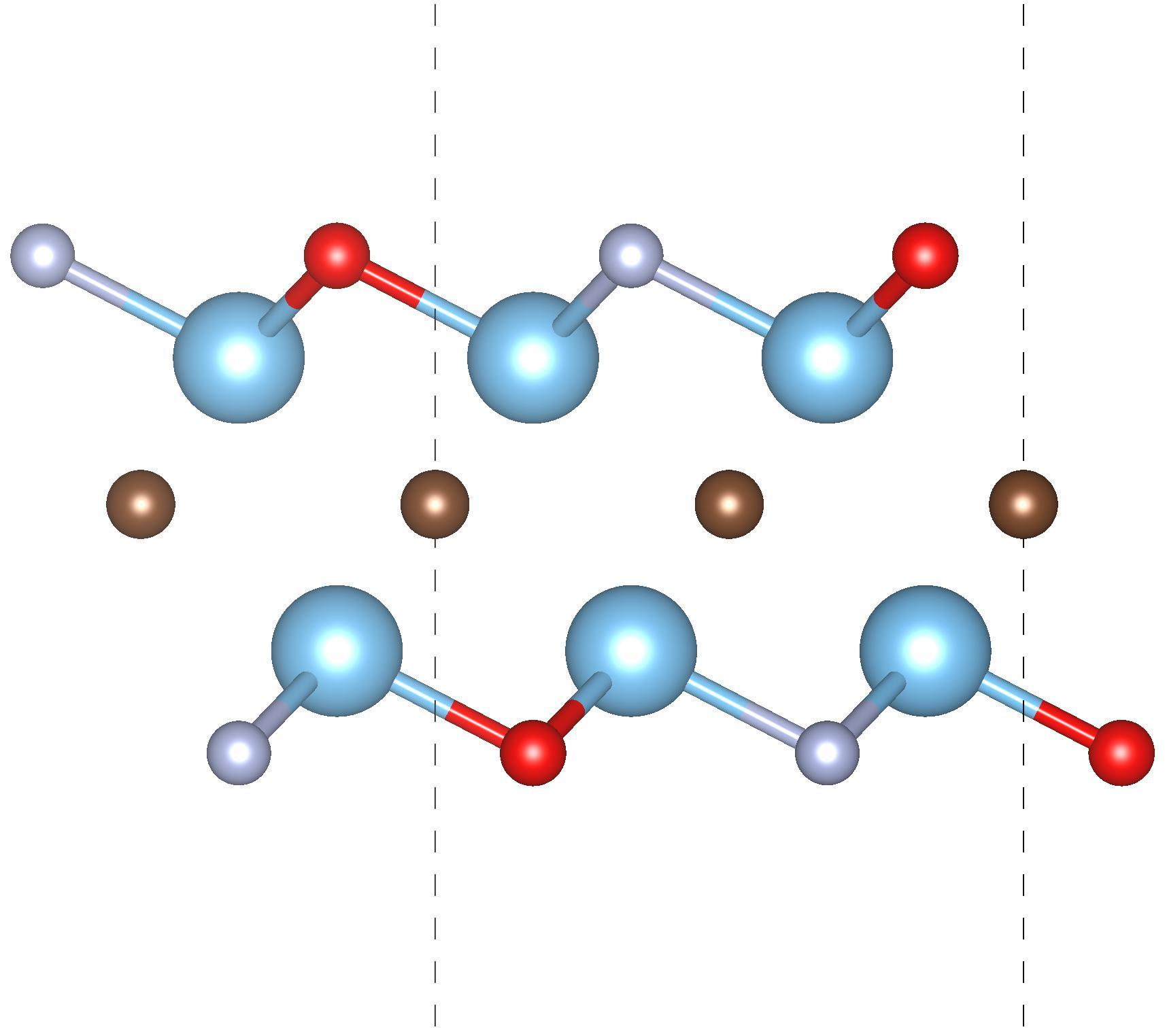, width=0.145\textwidth}} \quad
\subfigure[ ] {\epsfig{figure=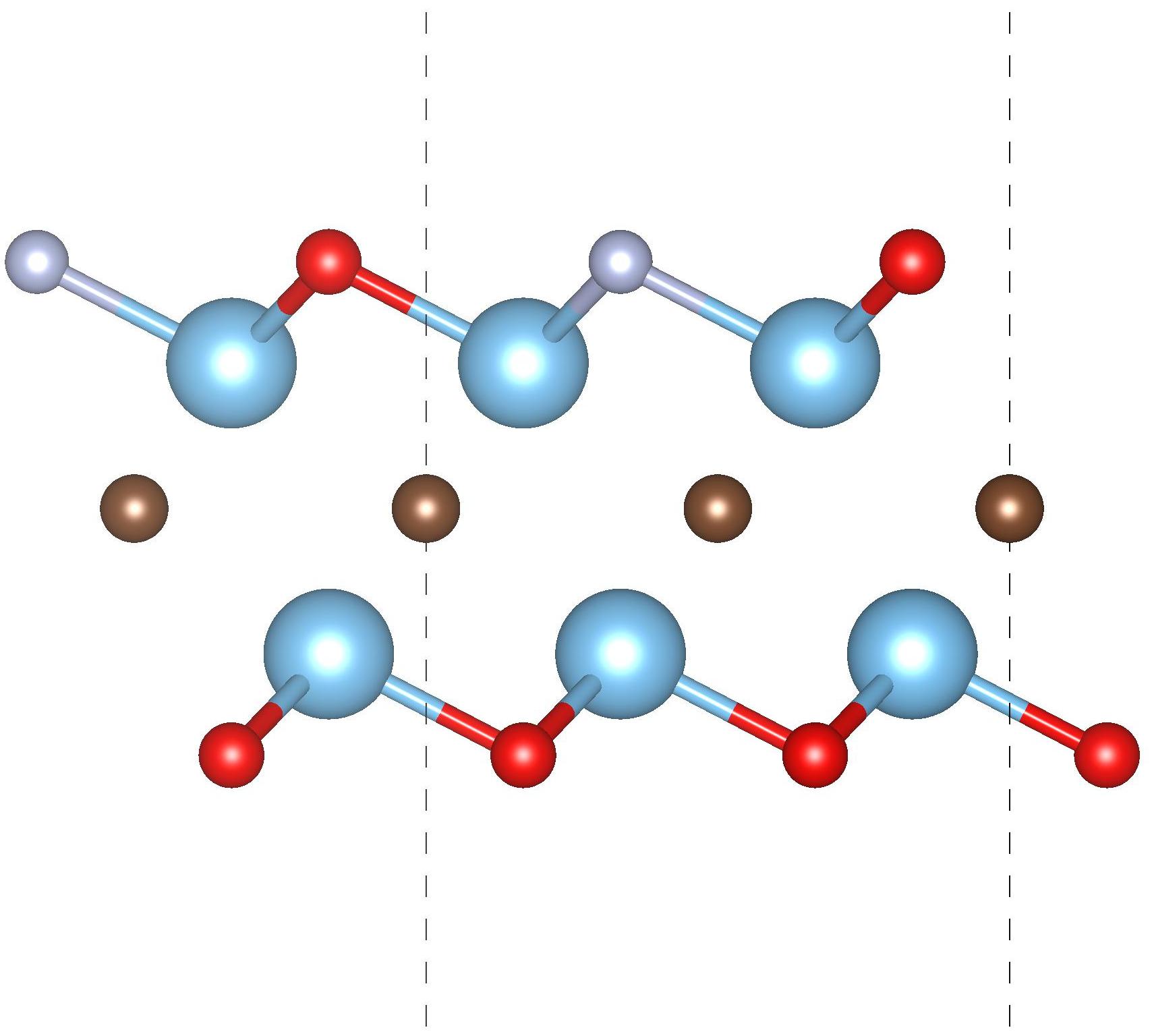, width=0.145\textwidth}} \quad
\subfigure[ ] {\epsfig{figure=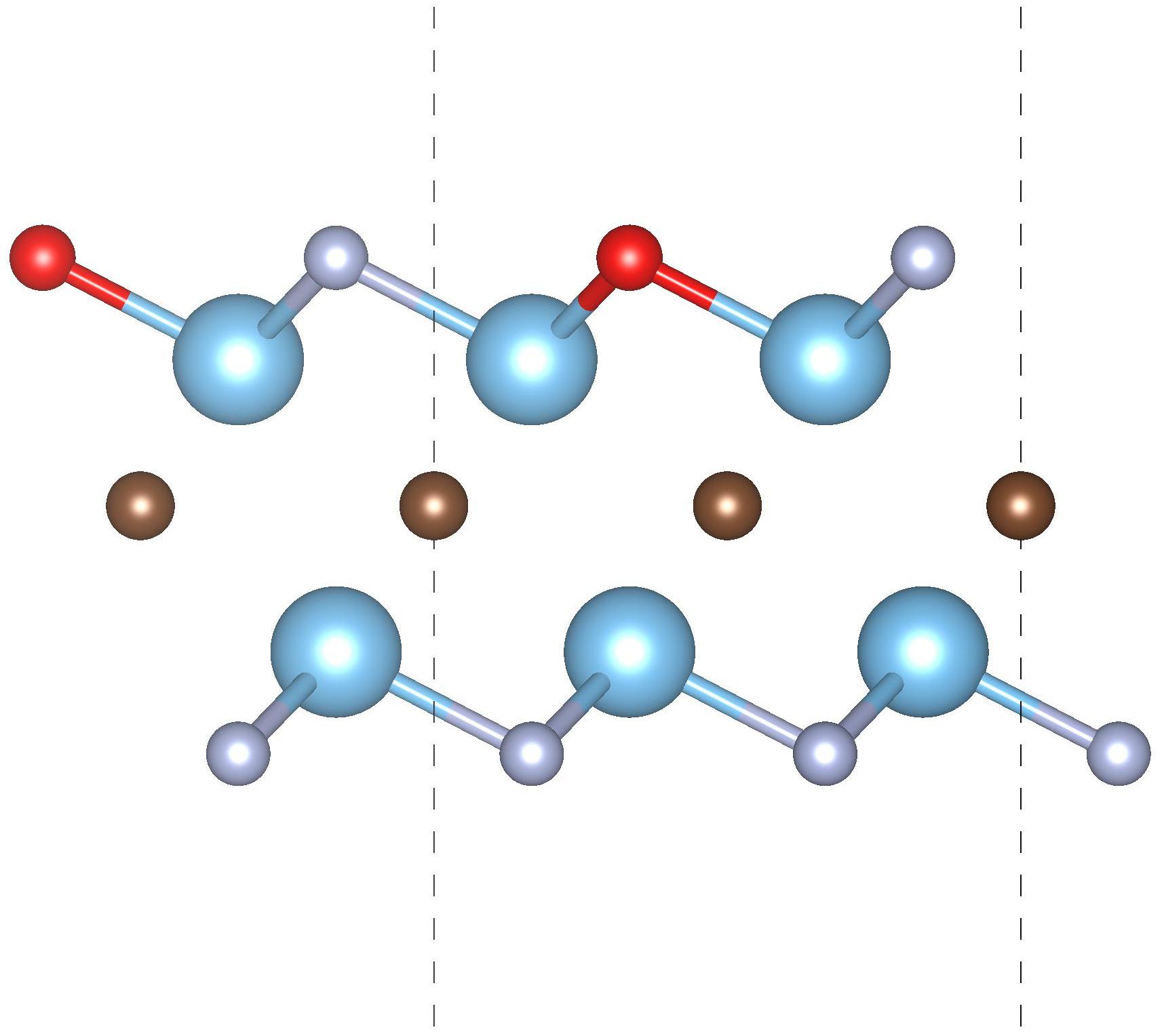, width=0.145\textwidth}}
\end{center}
\caption{Mixed functionalization in Ti$_{{\rm2}}$CA$_{{\rm x}}$B$_{{\rm y}}$ (A, B = O, F, OH). (a)Ti$_{{\rm2}}$CA$_{{\rm1.0}}$B$_{{\rm1.0}}$, (b) Ti$_{{\rm2}}$CA$_{{\rm1.5}}$B$_{{\rm0.5}}$, and (c) Ti$_{{\rm2}}$CA$_{{\rm0.5}}$B$_{{\rm1.5}}$. Blue, red, light gray and chocolate balls represent Ti, A, B and C atoms, respectively.}
\label{fig:mx-tic}
\end{figure}

\begin{table*}[htb]
\small
\caption{Calculated magnetic energy difference (E$_{{\rm FM,AFMj(j=1,2,3)}}$-E$_{{\rm FM}}$) and magnetic moments in of mixed Ti$_{{\rm2}}$CA$_{{\rm x}}$B$_{{\rm y}}$ (A, B = O, F, OH). The ground state energy is marked in bold. For comparison to DFT, the PBE results using Quantum Espresso code are listed in the table S1. The magnetic configurations here are starting guesses, then the system is relaxing to its self-consistent magnetic solution.} 
\centering
\begin{tabular*}{\textwidth}{@{\extracolsep{\fill}}lccccc||cccc}
\hline \hline
 & \multicolumn{4}{c}{Magnetic energy difference (meV)}&& \multicolumn{4}{c}{Total magnetic moment ($\mu_{B}$/cell)}\\
\cline{2-5}\cline{7-10}
Initial magnetic state & FM & AFM1 & AFM2 & AFM3 && FM & AFM1 & AFM2 & AFM3\\
\hline
Ti$_{{\rm2}}$CO$_{{\rm1}}$(OH)$_{{\rm1}}$     & {\bf0.00} & 0.00         & 36.30         & 36.30  &&  {\bf2.00} & 2.00      & 0.00 (AFM2)        & 0.00 (AFM2)\\
Ti$_{{\rm2}}$CO$_{{\rm1.5}}$(OH)$_{{\rm0.5}}$ & {\bf0.00} & 0.00         & 89.40         & 0.00   &&  {\bf1.00} & 1.00      & 0.40      & 1.00\\
Ti$_{{\rm2}}$CO$_{{\rm0.5}}$(OH)$_{{\rm1.5}}$ & {\bf0.00}      & 23.80   & 24.10         & 6.30   &&  {\bf2.31}      & 3.00 & 0.28      & 1.24\\ [0.45em]
Ti$_{{\rm2}}$CO$_{{\rm1}}$F$_{{\rm1}}$        & {\bf0.00} & 0.00         & 61.10         & 60.20  &&  {\bf2.00} & 2.00      & 0.00 (AFM2)        & 0.00 (AFM3)\\
Ti$_{{\rm2}}$CO$_{{\rm1.5}}$F$_{{\rm0.5}}$    & {\bf0.00} & 0.00         & 0.00          & 0.00   &&  {\bf1.00} & 1.00      & 1.00      & 1.00\\
Ti$_{{\rm2}}$CO$_{{\rm0.5}}$F$_{{\rm1.5}}$    & {\bf0.00} & 128.10       & 194.10        & 176.70 &&  {\bf3.00} & 0.96      & 0.06      & 2.21\\ [0.45em]
Ti$_{{\rm2}}$CF$_{{\rm1}}$(OH)$_{{\rm1}}$     & 0.00      & 200.20       & {\bf-29.60}   & 159.00 &&  4.00      & 0.48      & {\bf 0.00 (AFM2)}  & 0.00 (AFM3)\\
Ti$_{{\rm2}}$CF$_{{\rm1.5}}$(OH)$_{{\rm0.5}}$ & 0.00      & 44.20        & {\bf-165.40}  & 16.80  &&  3.34      & 0.17      & {\bf 0.00 (AFM2)}  & 0.06\\
Ti$_{{\rm2}}$CF$_{{\rm0.5}}$(OH)$_{{\rm1.5}}$ & {\bf0.00} & 263.00       & 148.20        & 108.30 &&  {\bf4.00} & 1.55      & 2.00      & 2.00\\
\hline
\end{tabular*} 
\label{enermagn-mix-DFT-DFTB} 
\end{table*}
To address the full range of mixed (O/OH, O/F and F/OH) surface termination compositions in the Ti$_{{\rm2}}$CA$_{{\rm x}}$B$_{{\rm y}}$ (A, B = O, F, OH and x, y=0.5, 1.0, 1.5 ; x+y = 2),  we used a 2$\times$1$\times$1-Ti$_{{\rm2}}$C supercell (i.e., Ti$_{{\rm4}}$C$_{{\rm2}}$A$_{{\rm a}}$B$_{{\rm b}}$, where a + b = 4). In Ti$_{{\rm2}}$CA$_{{\rm1.5}}$B$_{{\rm0.5}}$, Ti$_{{\rm2}}$CA$_{{\rm1.0}}$B$_{{\rm1.0}}$ and Ti$_{{\rm2}}$CA$_{{\rm0.5}}$B$_{{\rm1.5}}$ compositions, there are 3, 2 and 1 sites occupied by A termination and 1, 2 and 3 sites occupied by B termination, respectively. Surface composition is shown in Fig. \ref{fig:mx-tic}.   
\subsection{Electronic structure and magnetic properties of mixed Ti$_2$CA$_\mathrm{x}$B$_\mathrm{ y}$ MXene}
\begin{figure}[H] 
\centering
\includegraphics[width=0.4\textwidth]{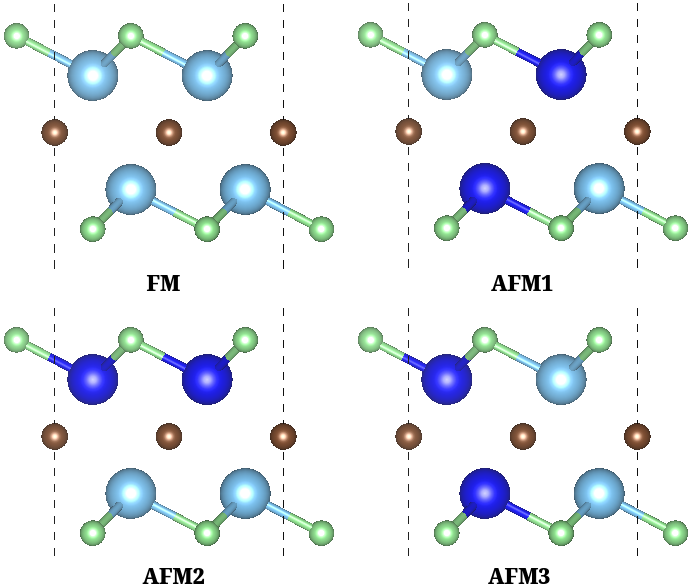}
\caption{Magnetic configurations of mixed Ti$_{{\rm2}}$CA$_{{\rm x}}$B$_{{\rm y}}$ MXene. Green and chocolate colors represent the terminal group (O, F, OH) and C atoms, respectively. The dark and light blue balls denote Ti atoms with opposite spin orientation.}
\label{magn-conf}
\end{figure}

\begin{figure}[H] 
\centering
\includegraphics[width=0.48\textwidth]{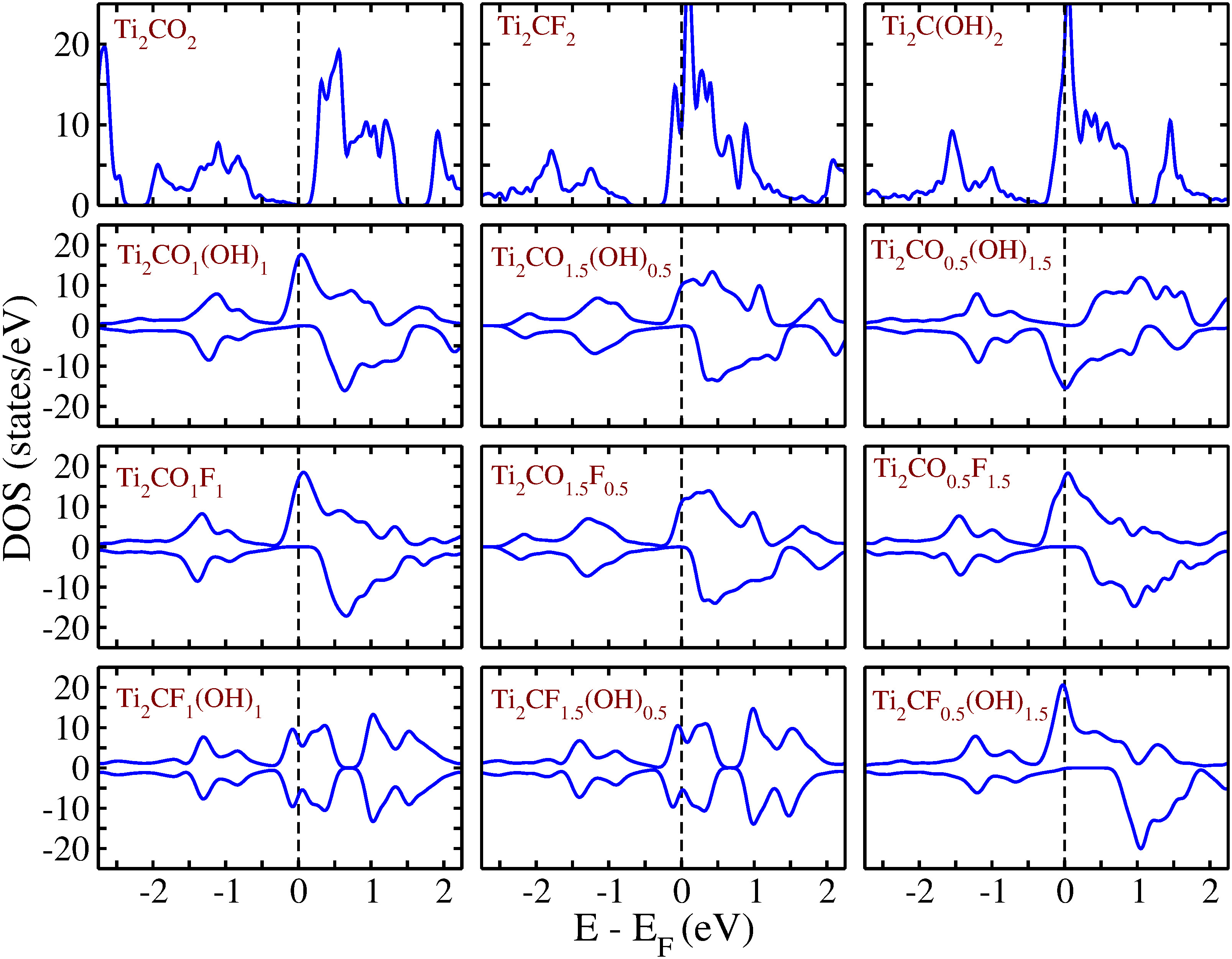}
\caption{Calculated total density of states (DOS) of bare Ti$_{{\rm2}}$CA$_{{\rm2}}$ and Ti$_{{\rm2}}$CA$_{{\rm x}}$B$_{{\rm y}}$ (A, B = O, F, OH) MXenes with different mixed terminations in their ground-state configurations. The Fermi level is set to zero. }
\label{dos-mix}
\end{figure}

\begin{figure}[H] 
\centering
\includegraphics[width=0.49\textwidth]{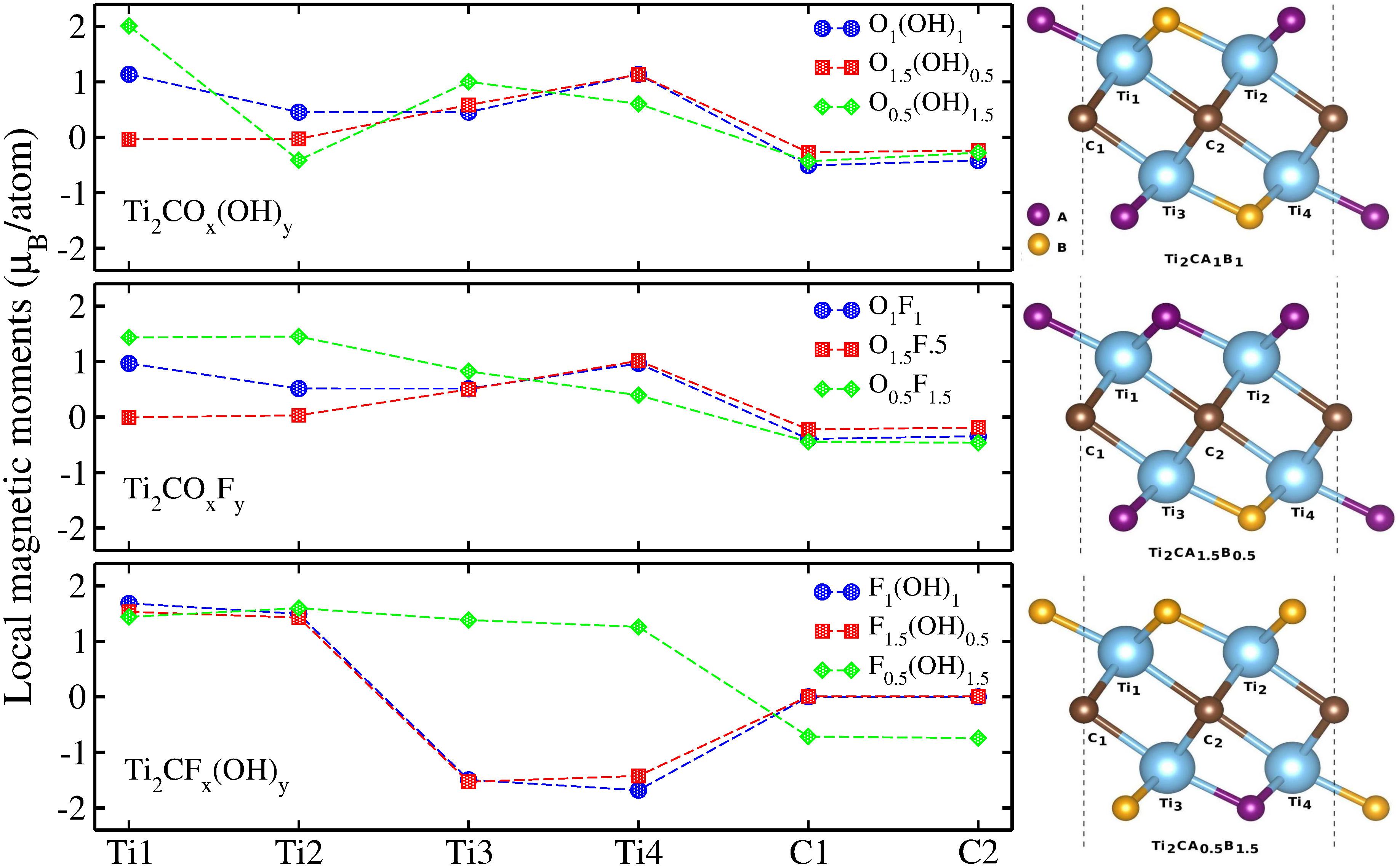}
\caption{Partial magnetic moment of different terminations in mixed Ti$_{{\rm2}}$CA$_{{\rm x}}$B$_{{\rm y}}$ (A, B = O, F, OH) MXene in their ground-state configurations, x, y=0.5, 1.0, 1.5. Dashed lines are guide to the eye.}
\label{part-mag-mix}
\end{figure}

In addition to the FM arrangement, three different AFM arrangements are considered here, including an interlayer antiferromagnetism (AFM2), and two possible intralayer antiferromagnetic arrangements (AFM1 and AFM3) as shown in Fig. \ref{magn-conf}. For AFM1, the Ti atoms on one side are in the same spin orientation with the diagonal Ti atoms on the other side, while in AFM2, the spin orientations of metal atoms on one side are opposite to those on the other side. In AFM3, the Ti atoms on one side are in the same spin orientation as the corresponding Ti atoms on the other side. Here, the calculations are based on the ground state of various configurations.
The electronic and magnetic properties of Ti$_{{\rm2}}$CA$_{{\rm x}}$B$_{{\rm y}}$ (A, B = O, F, OH) were explored. To determine the ground magnetic states of the considered composition of the terminal groups, the total energies of the possible magnetic alignments, including FM and AFM states, and their corresponding total magnetic moments were calculated and listed in Table \ref{enermagn-mix-DFT-DFTB}. We notice that the lowest energy has been found for FM state for most of the structures, except for Ti$_{{\rm2}}$CF$_{{\rm1}}$(OH)$_{{\rm1}}$ and Ti$_{{\rm2}}$CF$_{{\rm1.5}}$(OH)$_{{\rm0.5}}$, which prefer AFM2 alignment, where Ti atoms on one side are anti-ferromagnetically coupled with the Ti atoms on the other side. The energy differences between FM and AFM2 states in Ti$_{{\rm2}}$CF$_{{\rm1}}$(OH)$_{{\rm1}}$ and Ti$_{{\rm2}}$CF$_{{\rm1.5}}$(OH)$_{{\rm0.5}}$ are 29.6 and 165.4 meV/unit cell, respectively. We note that the  energetic difference is in the order of a few meV, this result is confirmed by DFT calculations using Quantum Espresso code (DFT-PBE results are listed in Table S1).\\
To understand the effects of surface termination composition and ordering on the electronic and magnetic properties of the studied systems, we plot the density of states for Ti$_{{\rm2}}$CA$_{{\rm x}}$B$_{{\rm y}}$ (A, B = O, F, OH) in Fig. \ref{dos-mix}. For all considered systems, regardless of the composition and ordering of O, F and OH terminations, Ti$_{{\rm2}}$CA$_{{\rm x}}$B$_{{\rm y}}$ exhibit nearly half-metallic ferromagnetism. There are many energy states crossing the Fermi level for the spin-up channels, indicating their metallic character, nonetheless, the spin-down channels are semiconducting. However, for Ti$_{{\rm2}}$CF$_{{\rm1}}$(OH)$_{{\rm1}}$ and Ti$_{{\rm2}}$CF$_{{\rm1.5}}$(OH)$_{{\rm0.5}}$, as shown in Fig. \ref{dos-mix}, both systems are metallic and the spin polarization is shown in an antiferromagnetic ordering, and thus they are AFM metals.  The anti-parallel spin alignment of the Ti atoms on one side and the Ti atoms on the other side results in the zero magnetic moment of the whole system and the AFM2 state in both Ti$_{{\rm2}}$CF$_{{\rm1}}$(OH)$_{{\rm1}}$ and Ti$_{{\rm2}}$CF$_{{\rm1.5}}$(OH)$_{{\rm0.5}}$. Our results reveal that the absolute value of the local magnetic moments of Ti atoms in FM-Ti$_{{\rm2}}$CF$_{{\rm0.5}}$(OH)$_{{\rm1.5}}$ are in the same range compared to those in AFM magnetic ground states of Ti$_{{\rm2}}$CF$_{{\rm1}}$(OH)$_{{\rm1}}$ and Ti$_{{\rm2}}$CF$_{{\rm1.5}}$(OH)$_{{\rm0.5}}$. Moreover, we found that the total magnetic moment values are sensitive to the terminal group composition. The total magnetic moments of the FM structures range from 1 $\mu_B$ in Ti$_{{\rm2}}$CO$_{{\rm1.5}}$(OH)$_{{\rm0.5}}$ to 4 $\mu_B$ in Ti$_{{\rm2}}$CF$_{{\rm0.5}}$(OH)$_{{\rm1.5}}$. In addition, Ti$_{{\rm2}}$CF$_{{\rm x}}$(OH)$_{{\rm y}}$ compositions possess the highest magnetic moments compared to Ti$_{{\rm2}}$CO$_{{\rm x}}$F$_{{\rm y}}$ and Ti$_{{\rm2}}$CO$_{{\rm x}}$(OH)$_{{\rm y}}$. Therefore, the varying concentration of O, F and OH provides an effective way to control the spin-polarization. Moreover, the magnetic moments of Ti$_{{\rm2}}$CA$_{{\rm x}}$B$_{{\rm y}}$ are mainly contributed by the Ti atoms and their values depend on the mixing ratio. 
Based on the ground-state configurations of the studied materials, the partial magnetic moments are present in Fig. \ref{part-mag-mix}. We clearly see that Ti$_{{\rm2}}$CO$_{{\rm1}}$(OH)$_{{\rm1}}$ and Ti$_{{\rm2}}$CO$_{{\rm1.5}}$(OH)$_{{\rm0.5}}$ represent similar trends as Ti$_{{\rm2}}$CO$_{{\rm1}}$F$_{{\rm1}}$ and Ti$_{{\rm2}}$CO$_{{\rm1.5}}$F$_{{\rm0.5}}$, respectively, for all ranges of compositions. In Ti$_{{\rm2}}$CO$_{{\rm1}}$(OH)$_{{\rm1}}$ and Ti$_{{\rm2}}$CO$_{{\rm1}}$F$_{{\rm1}}$, the magnetism is distributed over all Ti atoms with local magnetic moments of about 1, 0.5, 0.5, and 1 $\mu_B$, respectively. For Ti$_{{\rm2}}$CO$_{{\rm1.5}}$(OH)$_{{\rm0.5}}$ and Ti$_{{\rm2}}$CO$_{{\rm1.5}}$F$_{{\rm0.5}}$, a rapid increase of the spin moment going from Ti1 atom, where the local moment is almost zero, to Ti4 sites was seen. However, Ti$_{{\rm2}}$CO$_{{\rm0.5}}$(OH)$_{{\rm1.5}}$ and Ti$_{{\rm2}}$CO$_{{\rm0.5}}$F$_{{\rm1.5}}$ show different behavior. In the case of Ti$_{{\rm2}}$CF$_{{\rm x}}$(OH)$_{{\rm y}}$, both Ti$_{{\rm2}}$CF$_{{\rm1}}$(OH)$_{{\rm1}}$ and Ti$_{{\rm2}}$CF$_{{\rm1.5}}$(OH)$_{{\rm0.5}}$ show an AFM2 ordering where the local magnetic moments are 1.6 $\mu_B$ for Ti1 and Ti2 atoms and -1.6 $\mu_B$ for Ti3 and Ti4 atoms. However, a FM alignment was observed for the Ti$_{{\rm2}}$CF$_{{\rm0.5}}$(OH)$_{{\rm1.5}}$ composition, where the local magnetic moments of Ti1 and Ti2 atoms have almost the same value of about 1.5 $\mu_B$, while Ti3 and Ti4 atoms represent local magnetic moments of about 1.4 and 1.2 $\mu_B$, respectively. Noteworthy that the local moments at the C1 and C2 atoms are almost zero for all compositions.
\subsection{Substitution of one Ti by Sc in mixed surface termination MXene (Ti$_{{\rm2}}$CA$_{{\rm x}}$B$_{{\rm y}}$)}
To study the effect of the substitution of one Ti by Sc in mixed MXene, we used a 2$\times$2$\times$1-Ti$_{{\rm2}}$CA$_{{\rm x}}$B$_{{\rm y}}$ supercell (ScTi$_{{\rm15}}$C$_{{\rm8}}$A$_{{\rm a}}$B$_{{\rm b}}$, a+b=16; i.e., 40 to 52 atoms depending on the terminal group composition). 
After relaxing the structures, the total magnetic moments of 2$\times$2$\times$1-Ti$_{{\rm2}}$CA$_{{\rm x}}$B$_{{\rm y}}$ are calculated. The calculated total magnetic moments of ScTi$_{{\rm15}}$C$_{{\rm8}}$A$_{{\rm a}}$B$_{{\rm b}}$ (a+b=16) are listed in Table \ref{tab:mag-sc-mix} and the corresponding local magnetic moments are listed in Table S2 and Table S3. We see that the total magnetic moments are relatively higher for the mixed F and OH terminal group compared to mixed O and F and mixed O and OH terminations, ranging from 5.00$\mu_B$/cell for ScTi$_{{\rm15}}$C$_{{\rm8}}$F$_{{\rm12}}$(OH)$_{{\rm4}}$ to 15.06 $\mu_B$/cell for ScTi$_{{\rm15}}$C$_{{\rm8}}$F$_{{\rm4}}$(OH)$_{{\rm12}}$. The substitution of Sc atom in the mixed O and OH terminal group can induce lower magnetic moments of 7.04, 3.21 and 11.23 $\mu_B$/cell for ScTi$_{{\rm15}}$C$_{{\rm8}}$O$_{{\rm8}}$(OH)$_{{\rm8}}$, ScTi$_{{\rm15}}$C$_{{\rm8}}$O$_{{\rm12}}$(OH)$_{{\rm4}}$ and ScTi$_{{\rm15}}$C$_{{\rm8}}$O$_{{\rm4}}$(OH)$_{{\rm12}}$, respectively. For Sc substitution in 2$\times$2$\times$1-Ti$_{{\rm2}}$CO$_{{\rm x}}$F$_{{\rm y}}$ systems, the magnetic moments of ScTi$_{{\rm15}}$C$_{{\rm8}}$O$_{{\rm8}}$F$_{{\rm8}}$, ScTi$_{{\rm15}}$C$_{{\rm8}}$O$_{{\rm12}}$F$_{{\rm4}}$ and ScTi$_{{\rm15}}$C$_{{\rm8}}$O$_{{\rm4}}$F$_{{\rm12}}$ are 7.00, 3.00 and 11.00 $\mu_B$/cell, respectively. On the other hand, the introduction of Sc atom in the cell introduce a charge/spin redistribution keeping the magnetic moment in the same order as the structures without Ti substitution by Sc. The absolute value of magnetic moments of the carbon atoms slightly increased in all structures by the substitution (Please see Table S2 and Table S3 in the ESI). Remarkably, the introduction of one Sc in Ti$_{2}$CF$_{1.0}$(OH)$_{1.0}$ (i.e., ScTi$_{{\rm15}}$C$_{{\rm8}}$F$_{{\rm8}}$(OH)$_{{\rm8}}$) and Ti$_{2}$CF$_{1.5}$(OH)$_{0.5}$ (i.e., ScTi$_{{\rm15}}$C$_{{\rm8}}$F$_{{\rm4}}$(OH)$_{{\rm4}}$) makes a magnetic transition from AFM alignment to ferrimagnetic (FiM) ordering and the local magnetic moment were induced in the C atoms in the order of 0.6 to 0.75$\mu_B$. Moreover, the value of the local magnetic moments on Sc atoms is slightly low (in the order of 0.14 and 0.18 $\mu_B$).
\begin{table}[H]
\caption{Calculated total magnetic moments of 
2$\times$2$\times$1-Ti$_{{\rm2}}$CA$_{{\rm x}}$B$_{{\rm y}}$ (A, B = O, F, OH) with one Ti substituted by Sc.}
\centering
\begin{tabular*}{0.48\textwidth}{@{\extracolsep{\fill}}lc}
\hline \hline
Structure & Magnetic moment ($\mu_{B}$/cell)\\
\hline
ScTi$_{{\rm15}}$C$_{{\rm8}}$O$_{{\rm8}}$(OH)$_{{\rm8}}$  & 7.04  \\ 
ScTi$_{{\rm15}}$C$_{{\rm8}}$O$_{{\rm12}}$(OH)$_{{\rm4}}$ & 3.21  \\ 
ScTi$_{{\rm15}}$C$_{{\rm8}}$O$_{{\rm4}}$(OH)$_{{\rm12}}$ & 11.23 \\  [0.45em] 
ScTi$_{{\rm15}}$C$_{{\rm8}}$O$_{{\rm8}}$F$_{{\rm8}}$     & 7.00  \\ 
ScTi$_{{\rm15}}$C$_{{\rm8}}$O$_{{\rm12}}$F$_{{\rm4}}$    & 3.00  \\ 
ScTi$_{{\rm15}}$C$_{{\rm8}}$O$_{{\rm4}}$F$_{{\rm12}}$    & 11.00 \\ [0.45em] 
ScTi$_{{\rm15}}$C$_{{\rm8}}$F$_{{\rm8}}$(OH)$_{{\rm8}}$  & 14.97 \\ 
ScTi$_{{\rm15}}$C$_{{\rm8}}$F$_{{\rm12}}$(OH)$_{{\rm4}}$ & 5.00  \\ 
ScTi$_{{\rm15}}$C$_{{\rm8}}$F$_{{\rm4}}$(OH)$_{{\rm12}}$ & 15.06 \\ 
\hline
\end{tabular*} \label{tab:mag-sc-mix}
\end{table}
\subsection{Substitution of Ti by Sc in Ti$_{{\rm2}}$CO$_{{\rm2}}$}
In our previous study \cite{monACSOmega}, we show that the presence of C-vacancy in Ti$_{{\rm2}}$CO$_{{\rm2}}$ induces magnetism in the system. Moreover, as we noticed above, there is some effect of Sc atom in Ti$_{{\rm2}}$C-based MXene structures (cf. Section 3.1 and 3.2). We therefore decided to investigate also the effect of Sc substitution in a bare Ti$_{{\rm2}}$CO$_{{\rm2}}$ MXene. Here, we investigate the effect of the substitution of the transition metal (Ti) by Sc in 5$\times$5$\times$1-Ti$_{{\rm2}}$CO$_{{\rm2}}$; i.e., Sc$_{{\rm n}}$Ti$_{{\rm50-n}}$C$_{{\rm25}}$O$_{{\rm50}}$ structure, where n = 1, 2, 3, 4, 5, 6, and 7. 
\begin{table}[H]
\caption{Calculated magnetic moments (in $\mu_{B}$/cell) of Ti substituted by Sc in 5$\times$5$\times$1-Ti$_{{\rm2}}$CO$_{{\rm2}}$ on different directions (Zigzag, Armchair and Cluster). NM refers to nonmagnetic.}
\centering
\begin{tabular*}{0.48\textwidth}{@{\extracolsep{\fill}}lccc}
\hline \hline
  & Zigzag & Armchair & Cluster \\
\hline
1Sc & 0.00 (NM)   & 0.00 (NM) & 0.00 (NM) \\
2Sc & 0.16                         & 0.16                       & 0.16 \\ 
3Sc & 0.18                         & 0.18                       & 0.18 \\
4Sc & 0.00 (NM)   & 0.00 (NM) & 0.00 (NM) \\
5Sc & 1.05                         & 0.00 (NM) & 0.96 \\
6Sc & 2.08                         & 0.00 (NM) & 1.98 \\
7Sc & 3.07                         & 1.78                       & -- \\
\hline
\end{tabular*} 
\label{tab:mag-sc-t2co2}
\end{table}
In the Table \ref{tab:mag-sc-t2co2}, we show the calculated magnetic moments in Ti$_{{\rm2}}$CO$_{{\rm2}}$ supercell after the substitution of n Ti atom by n Sc atoms. In our study, we have found that by substituting only one Ti by Sc, the structure remain nonmagnetic (NM). 
By increasing the number of Sc atoms in the supercell, the magnetism is turned on, but some structures keep a nonmagnetic (NM) behavior, more precisely, when substituting 4 Ti atoms with Sc, and also when substituting 5 and 6 Ti atoms with Sc along the armchair direction. Substitution induces local magnetic moments from 0.16$\mu_B$ to about 3$\mu_B$ in the supercell depending on the position and the number of Sc in the system.\\
We note that for the magnetic structure, 5 and 6 Sc in 5$\times$5$\times$1-Ti$_{{\rm2}}$CO$_{{\rm2}}$ in the zigzag direction are more preferred over the cluster form by an energy difference of 0.08 eV and 0.24 eV, respectively. While 7Sc in armchair are more stable compared to the zigzag direction, the energy difference is 0.32 eV. The total energies of n Ti atoms substituted by n Sc atoms in 5$\times$5$\times$1-Ti$_{{\rm2}}$CO$_{{\rm2}}$ following the Zigzag, Armchair, and Cluster directions are listed in Table S2.
\begin{figure}[H]
{\centering
\captionsetup{justification=centering} 
\subfigure[ ] {\epsfig{figure=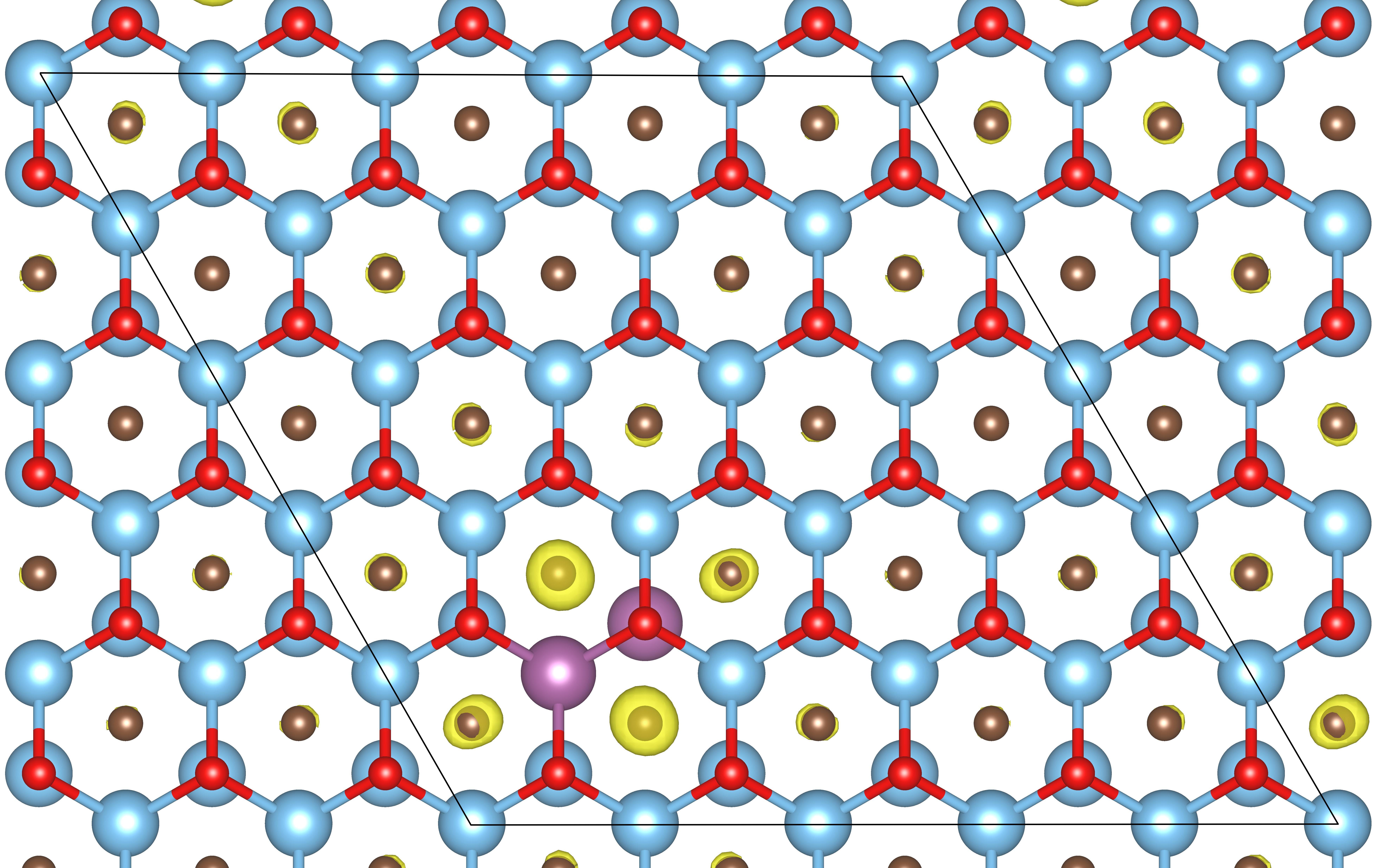, width=0.23\textwidth}} \quad
\subfigure[ ] {\epsfig{figure=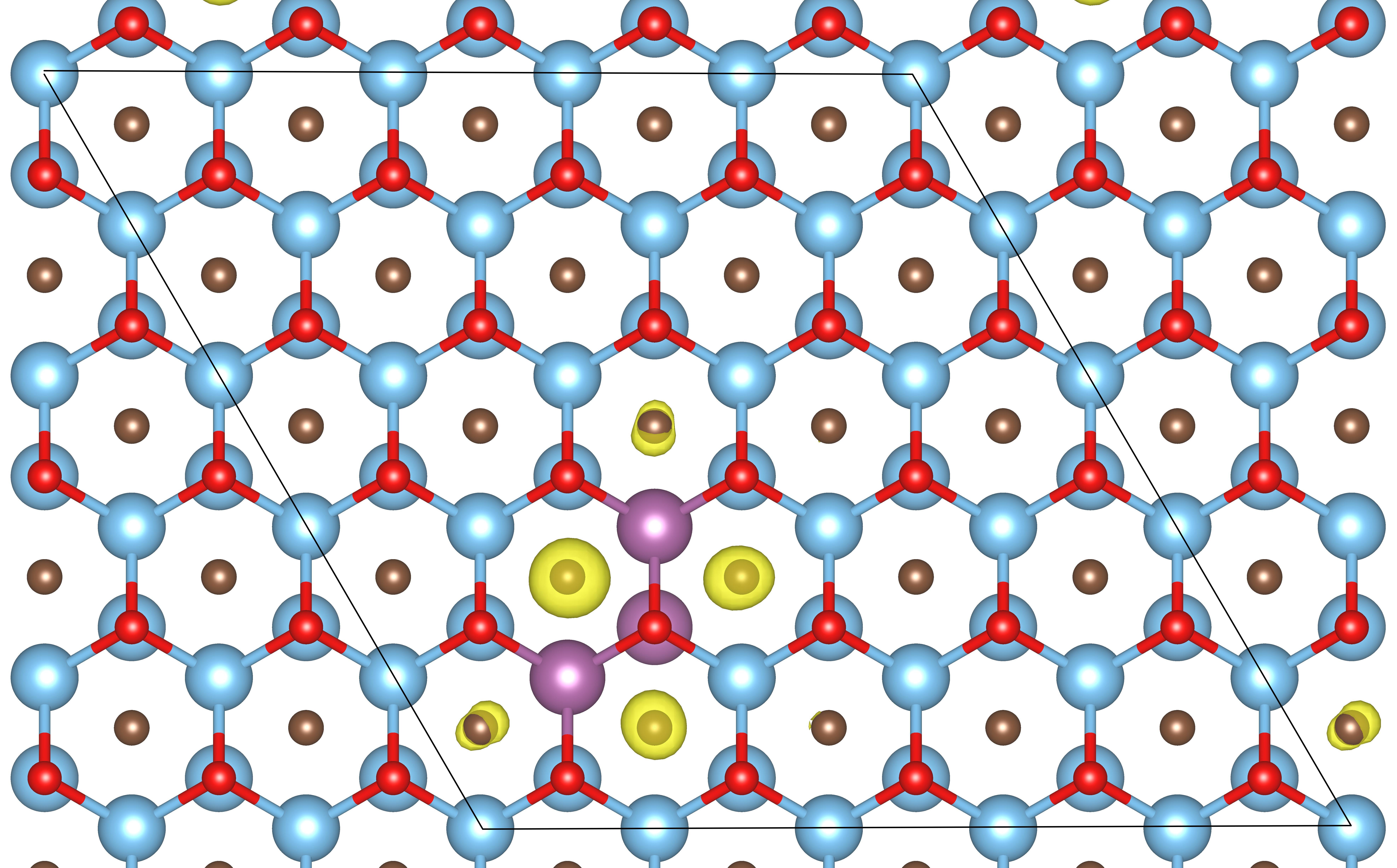, width=0.23\textwidth}} \\
\subfigure[ ] {\epsfig{figure=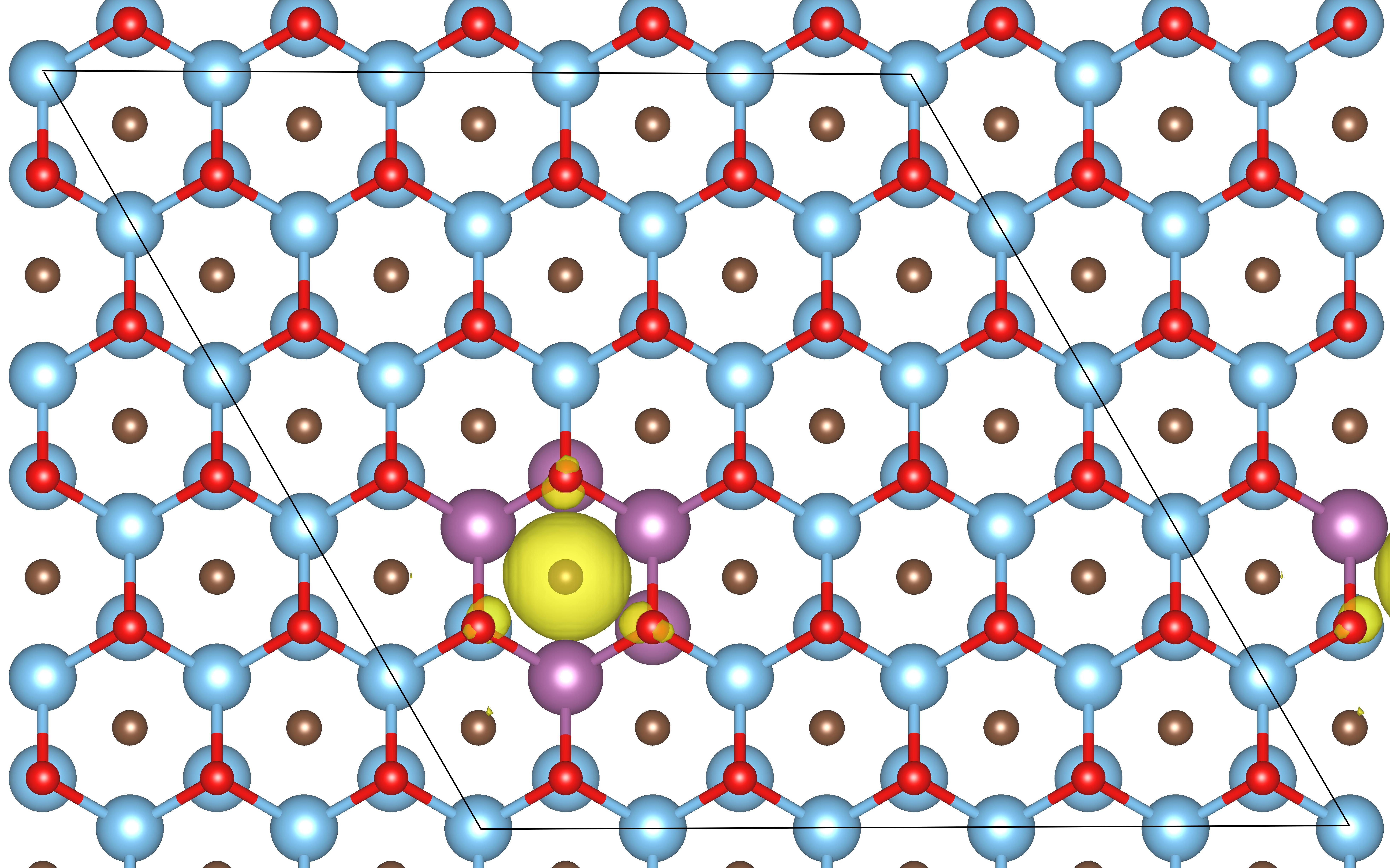, width=0.23\textwidth}} \quad
\subfigure[ ] {\epsfig{figure=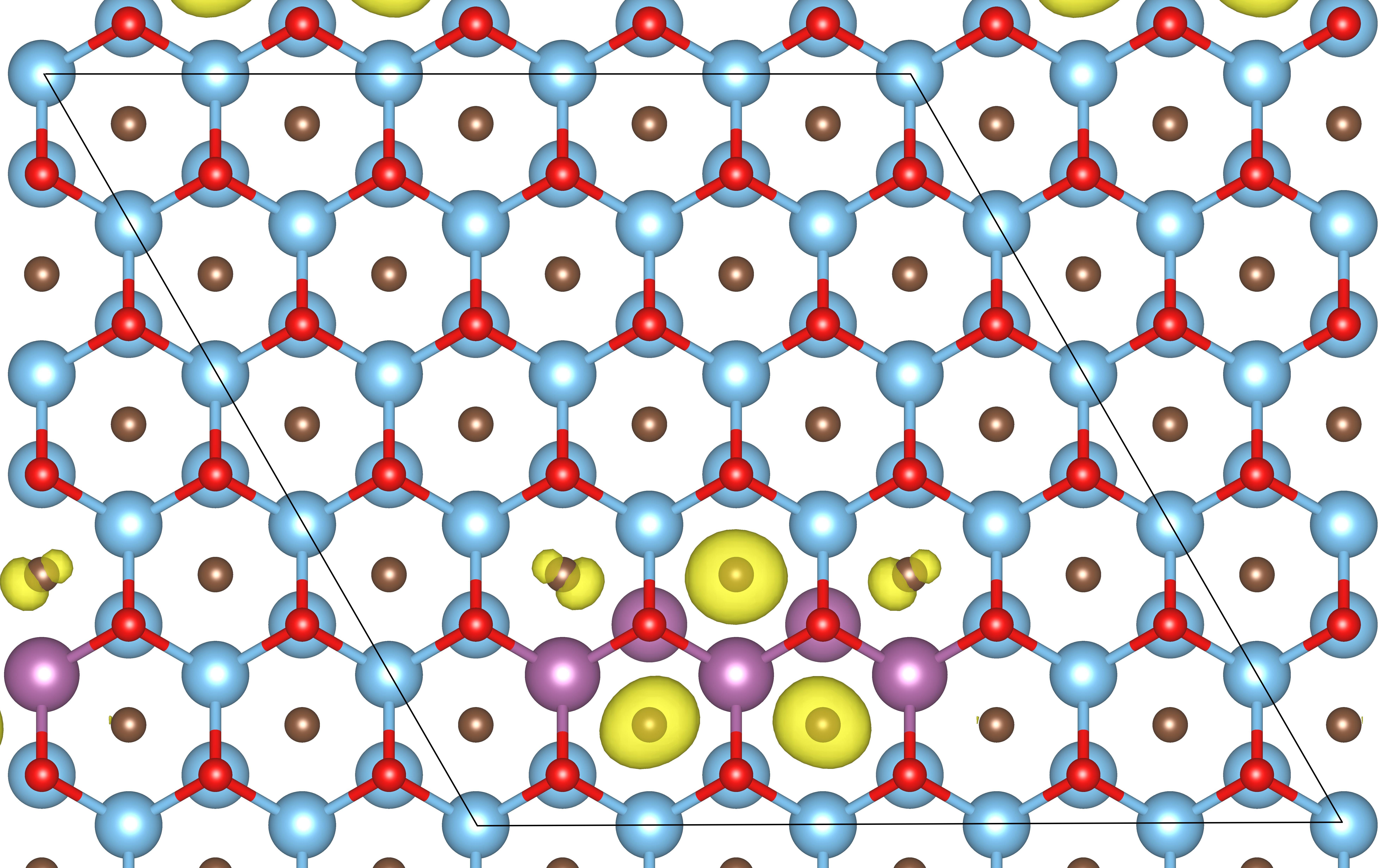, width=0.23\textwidth}} \\
\subfigure[ ] {\epsfig{figure=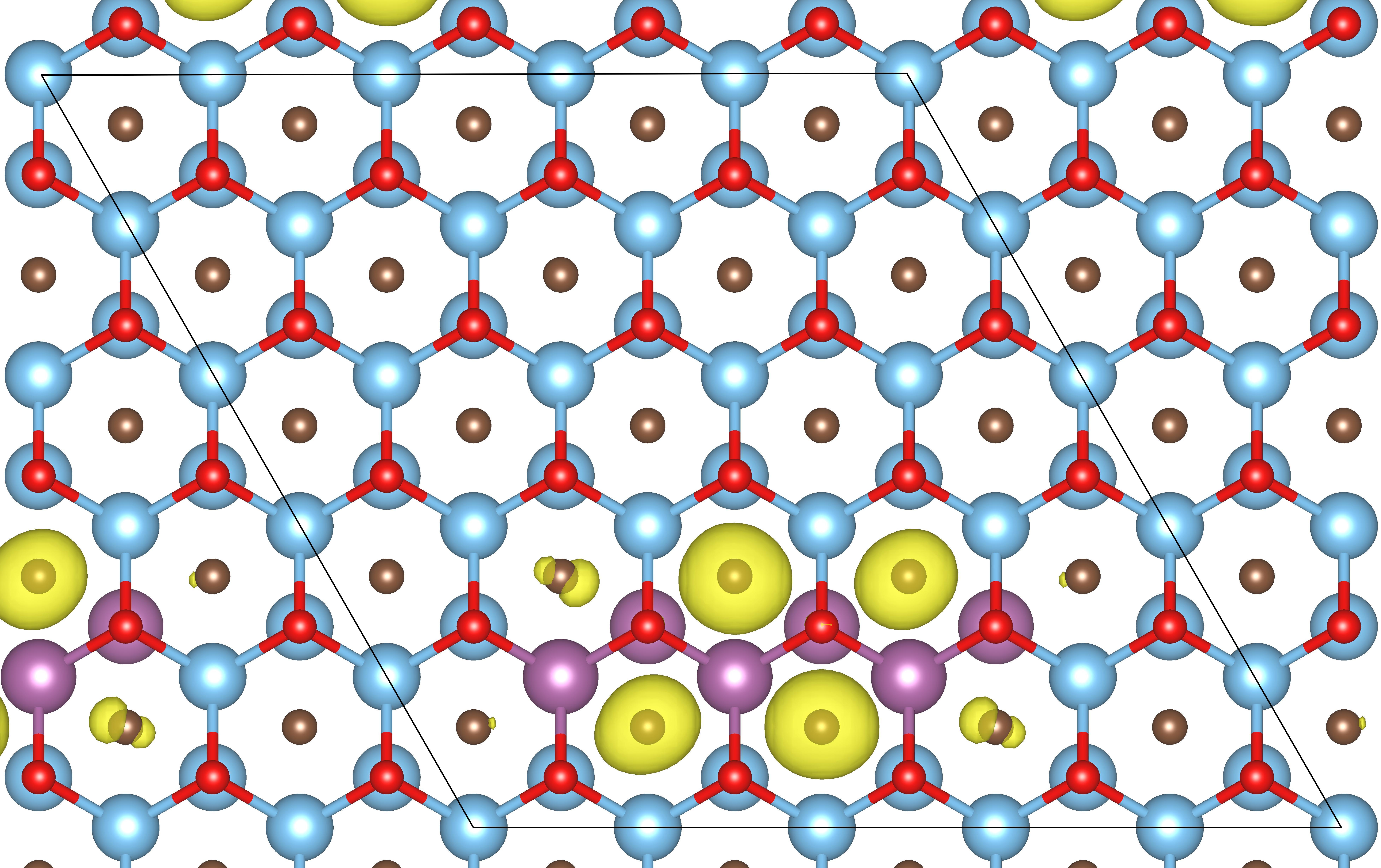, width=0.23\textwidth}} \quad
\subfigure[ ] {\epsfig{figure=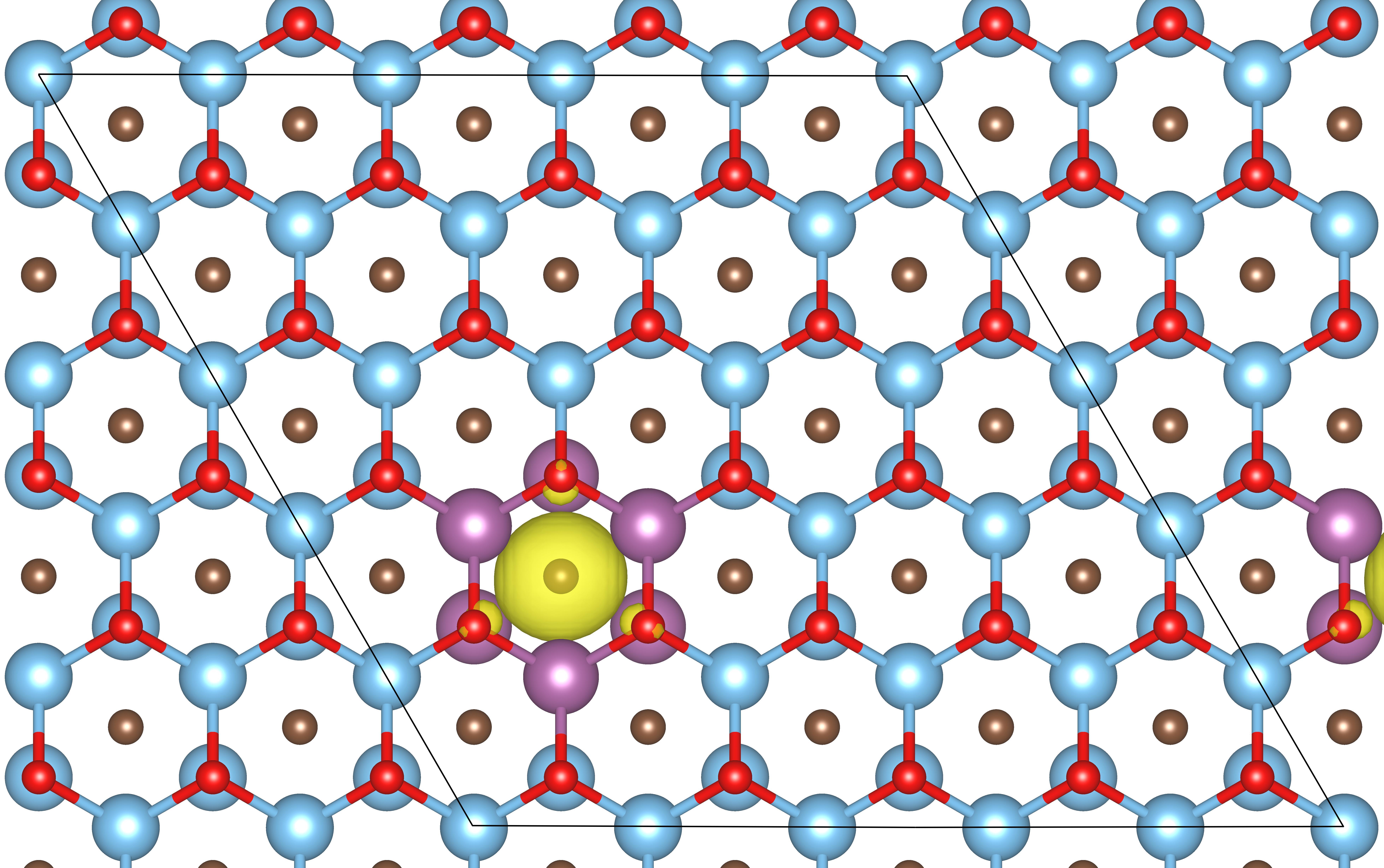, width=0.23\textwidth}}\\
\subfigure[ ] {\epsfig{figure=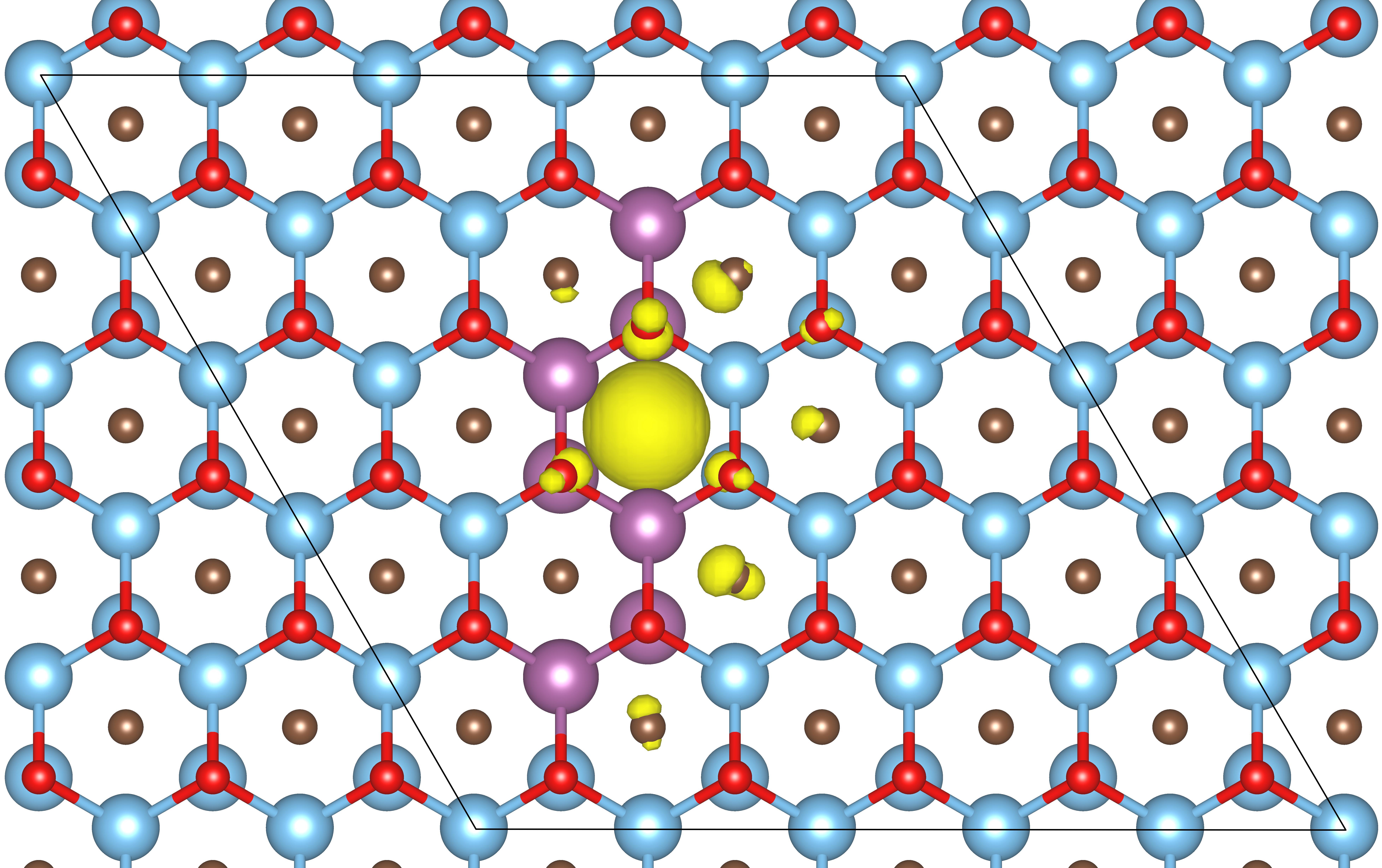, width=0.23\textwidth}} \quad
\subfigure[ ] {\epsfig{figure=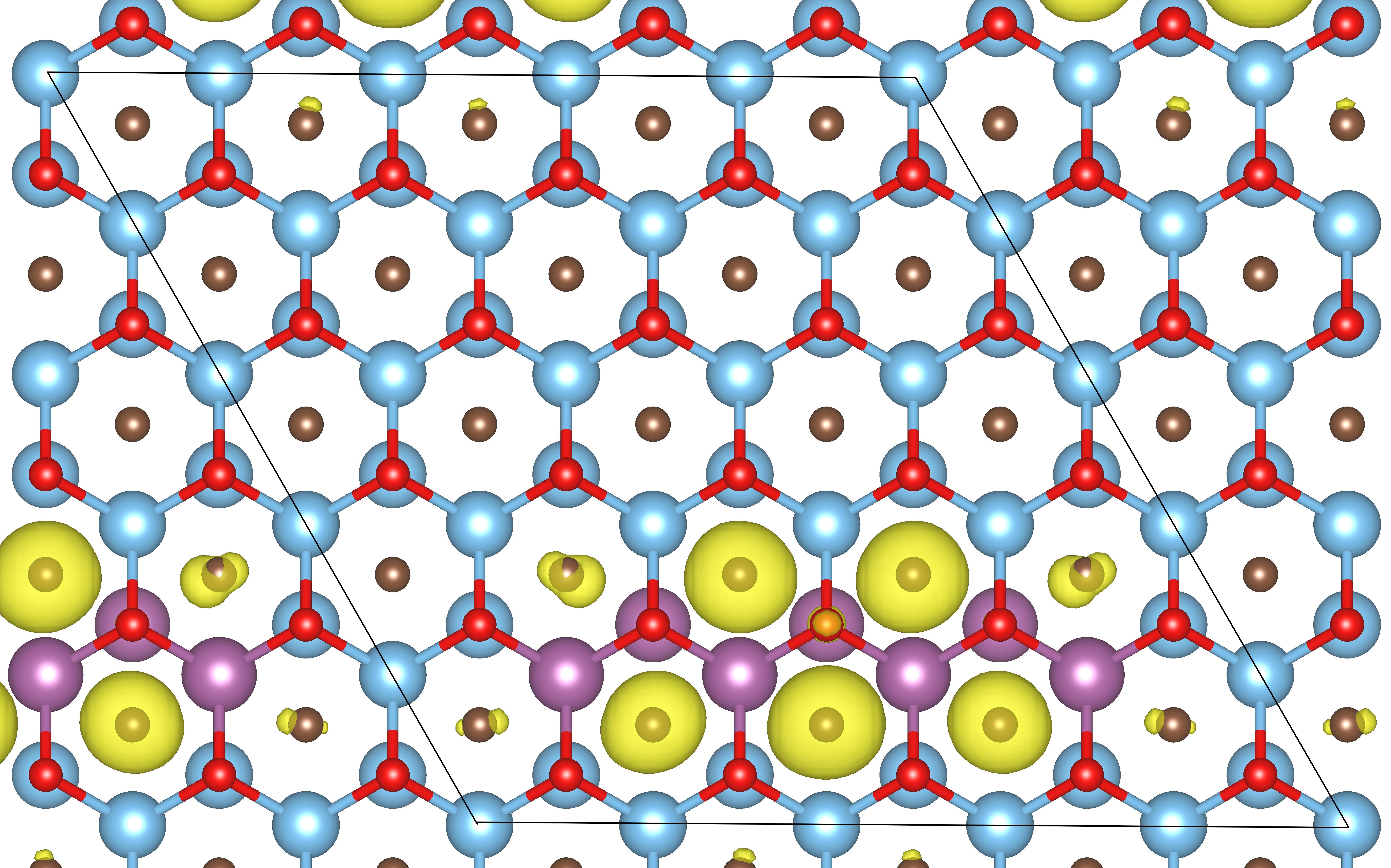, width=0.23\textwidth}}  }
\caption{Spin density distribution of n Ti atoms substituted by n Sc atoms in 5$\times$5$\times$1-Ti$_{{\rm2}}$CO$_{{\rm2}}$, n=2, 3, 5, 6, and 7. (a) 2 Sc atoms and (b) 3 Sc atoms using isosurface =10$^{{\rm -3}}$ au and (c) 5 Sc atoms (cluster) (d) 5 Sc atoms (zigzag direction) (e) 6 Sc atoms (zigzag direction) (f) 6 Sc atoms (cluster) (g) 7 Sc atoms (armchair direction), and (h) 7 Sc atoms (zigzag direction) with isosurface = 5$\times$10$^{{\rm -3}}$ au. Yellow color represents the spin density. Blue, red, chocolate and violet balls refer to Ti, O, C and Sc atoms, respectively. The plots were produced by using VESTA software\cite{vesta}. }
\label{fig:sp-tm-on-tico}
\end{figure}
To pick up more knowledge into the magnetic properties, the spin electron density distributions of n Sc atoms in Ti$_{{\rm50}}$C$_{{\rm25}}$O$_{{\rm50}}$ are computed and compared in Fig. \ref{fig:sp-tm-on-tico}. The results indicate that the spin density is almost entirely localized around C atoms that are close to the Sc atoms. There is no spin density in the rest of the supercell, which confirm that the incorporation of the Sc atom make a charge rearrangement around the surrounding atoms and therefore, magnetism is induced. It is shown that the induced magnetism is mainly attributed the Sc since the spin electrons are mostly concentrated on the surrounding of Sc atoms. Moreover, some fraction of spin density are located at C atoms. 
\section{Conclusion}
The electronic structure and magnetic properties of mixed termination in Ti$_{{\rm2}}$C MXenes have been studied by DFTB method using the recently developed xTB Hamiltonian. Ti$_{{\rm2}}$CA$_{{\rm x}}$B$_{{\rm y}}$ (A, B = O, F, OH) monolayers show spin polarization and the magnetic alignment depends on the mixed termination. Ferromagnetic (FM) ordering appears in most of compositions except Ti$_{{\rm2}}$CF$_{{\rm1}}$(OH)$_{{\rm1}}$ and Ti$_{{\rm2}}$CF$_{{\rm1.5}}$(OH)$_{{\rm0.5}}$, which are antiferromagnetic (spin orientations of metal atoms on one side are opposite to those on the other side; AFM2 arrangement). The effect of the transition metal atom (Ti) substituted by the another one (Sc atom) on the electronic and magnetic properties were modeled by 2$\times$2$\times$1-Ti$_{{\rm2}}$CA$_{{\rm x}}$B$_{{\rm y}}$ supercell. A ferrimagnetic (FiM) alignment was found and a charge/spin redistribution was introduced which explain the slight change of the total magnetic moments compared to the structures without Sc atoms. Interestingly, a magnetic transition from AFM to FiM was found for ScTi$_{{\rm15}}$C$_{{\rm8}}$F$_{{\rm8}}$(OH)$_{{\rm8}}$ and ScTi$_{{\rm15}}$C$_{{\rm8}}$F$_{{\rm12}}$(OH)$_{{\rm4}}$. We also investigated the effect of substitution of n Ti atoms by n Sc atoms in the fully O-terminated Ti$_{{\rm2}}$CO$_{{\rm2}}$ MXene following different configuration, including cluster, armchair, and zigzag direction. We found that magnetism is tuned by the Sc substitution in Ti$_{{\rm50}}$C$_{{\rm25}}$O$_{{\rm50}}$ and it is sensitive to the number of Sc atoms in the cell and to the direction of the substitutent. In the magnetic structures, the total magnetic moment is higher for higher number of Sc in the cell. Moreover, the shown half-metallicity ferromagnetism and metallicity antiferromagnetism in Ti$_{{\rm2}}$CA$_{{\rm x}}$B$_{{\rm y}}$ and Ti$_{{\rm2}}$CO$_{{\rm 2}}$ may provide more potential applications in spintronic devices.\\
Finally, we conclude that the recent GFN1-xTB Hamiltonian is useful for the investigation of the electronic and magnetic properties of MXenes. However, further targeted studies are required to conclude about it’s utilization for the modeling of the 2D materials and periodic systems of big sizes. 

\section*{Conflicts of interest}
There is no conflict to declare.
\section*{Acknowledgements}
This article has been produced with the financial support of the Czech Science Foundation (21-28709S) and the European Union under the LERCO project (number CZ.10.03.01/00/22$\_$003/0000003) via the Operational Programme Just Transition. The computations were performed at IT4Innovations National Supercomputing Center through the e-INFRA CZ (ID:90140).





\bibliography{refs} 
\bibliographystyle{rsc} 

\end{document}